\documentclass[twocolumn,aip,reprint,numerical]{revtex4-1}
\usepackage[latin9]{inputenc}
\usepackage{amsmath}
\usepackage{amssymb}
\usepackage{graphicx}

\makeatletter
\usepackage{dcolumn}
\usepackage{bm}
\usepackage{epsfig}
\usepackage{braket}

\bibpunct{(}{)}{,}{n}{}{}

\setcitestyle{round}
\renewcommand{\fnum@figure}{\textbf{Fig.~\thefigure}}
\usepackage{upgreek}

\makeatother

\begin{document}
\title{Observation of persistent orientation of chiral molecules by laser
field with twisted polarization}
\author{Ilia Tutunnikov}
\affiliation{AMOS and Department of Chemical and Biological Physics, The Weizmann
Institute of Science, Rehovot 7610001, Israel}
\author{Johannes Flo{\ss}}
\affiliation{Chemical Physics Theory Group, Department of Chemistry, and Center
for Quantum Information and Quantum Control, University of Toronto,
Toronto, ON M5S 3H6, Canada}
\author{Erez Gershnabel}
\affiliation{AMOS and Department of Chemical and Biological Physics, The Weizmann
Institute of Science, Rehovot 7610001, Israel}
\author{Paul Brumer}
\affiliation{Chemical Physics Theory Group, Department of Chemistry, and Center
for Quantum Information and Quantum Control, University of Toronto,
Toronto, ON M5S 3H6, Canada}
\author{Ilya Sh. Averbukh}
\email{ilya.averbukh@weizmann.ac.il}

\affiliation{AMOS and Department of Chemical and Biological Physics, The Weizmann
Institute of Science, Rehovot 7610001, Israel}
\author{Alexander A. Milner}
\affiliation{Department of Physics \& Astronomy, The University of British Columbia,
Vancouver, BC V6T 1Z1, Canada}
\author{Valery Milner}
\email{vmilner@phas.ubc.ca}

\affiliation{Department of Physics \& Astronomy, The University of British Columbia,
Vancouver, BC V6T 1Z1, Canada}
\begin{abstract}
Molecular chirality is an omnipresent phenomenon of fundamental significance
in physics, chemistry and biology. For this reason, search for novel
techniques for enantioselective control, detection and separation
of chiral molecules is of particular importance. It has been recently
predicted that laser fields with twisted polarization may induce persistent
enantioselective field-free orientation of chiral molecules. Here
we report the first experimental observation of this phenomenon using
propylene oxide molecules ($\mathrm{CH_{3}CHCH_{2}O}$, or PPO) spun
by an optical centrifuge - a laser pulse, whose linear polarization
undergoes an accelerated rotation around its propagation direction.
We show that PPO molecules remain oriented on a time scale exceeding
the duration of the centrifuge pulse by several orders of magnitude.
The demonstrated long-time field-free enantioselective orientation
opens new avenues for optical manipulation, discrimination, and, potentially,
separation of molecular enantiomers.
\end{abstract}
\maketitle
Chiral molecules exist in two nonsuperimposable forms called enantiomers
\cite{Cotton}. The ability to analyze and separate mixtures of enantiomers
is crucial, for example, in drug synthesis as different enantiomers
of chiral molecules may exhibit strikingly different biological activity.
Over the years, various optical approaches have been developed to
control molecules in the gas phase and induce their alignment/orientation
(for recent reviews, see e.g. \cite{Koch2018,Lemeshko2013,Fleischer2012,Ohshima2010}).
The most known technique of this kind is field-free \textit{alignment}
of linear molecules by short linearly polarized laser pulses, which
makes the most polarizable molecular axis point along the polarization
direction after the end of the pulse \cite{Stapelfeldt2003}. Laser-induced
molecular \textit{orientation} is a more challenging task; however,
several techniques have been suggested and demonstrated for impulsive
orientation of linear and even asymmetric top molecules under field-free
condition (see, e.g. \cite{Lin2018}, and references therein).

Owing to the symmetry of light interaction with the induced dipole
moment, strong laser fields with fixed linear polarization can be
utilized only to align, but not orient, molecules in space. In contrast,
fields with twisting polarization break this symmetry, and were predicted
to transiently orient chiral molecules \cite{Yachmenev2016,Gershnabel2018,Tutunnikov2018},
as was recently confirmed experimentally \cite{Milner2019}.

Here we report a related, yet fundamentally new phenomenon -- field-free
\emph{persistent} enantioselective orientation (PESO) of chiral molecules,
which lasts orders of magnitude longer than the duration of the exciting
laser pulses. This effect is unique for chiral molecules driven by
nonresonant laser fields with twisted polarization. In all the previously
known approaches to impulsive molecular orientation, including techniques
using single-cycle THz pulses \cite{Harde1991,Dion1999,Averbukh2001,Machholm2001,Fleischer2011,Kitano2013,Damari2016},
alone or in combination with optical pulses \cite{Daems2005,Gershnabel2006,Egodapitiya2014},
or two-color laser fields \cite{Vrakking1997,Dion1999,Kanai2001,De2009,Oda2010,JW2010,Frumker2012,Takemoto2008,Lin2018},
the post-pulse orientation shows transient signals with zero time
average. To the best of our knowledge, our paper presents the first
demonstration of the long-lasting permanent molecular orientation
induced by a pulsed optical field.

Long-lived orientation of chiral molecules induced by laser fields
with twisted polarization was recently theoretically predicted in
\cite{Gershnabel2018,Tutunnikov2018,Tutunnikov2019}. Several possible
examples of such fields include a pair of delayed cross-polarized
laser pulses \cite{Fleischer2009,Kitano2009}, chiral pulse trains
\cite{Zhdanovich2011,Bloomquist2012,Floss2012}, polarization-shaped
pulses \cite{Karras2015,Prost2017,Prost2018} and an optical centrifuge
\cite{Karczmarek1999,Villeneuve2000,Yuan2011,Korobenko2014-PRL,Korobenko2018}.

In what follows, we present the first experimental observation of
the PESO phenomenon, using chiral propylene oxide molecules spun by
an optical centrifuge. The results of experimental measurements are
in a good agreement with the theoretical predictions.
\begin{figure*}
\noindent \centering{}\includegraphics[scale=0.75]{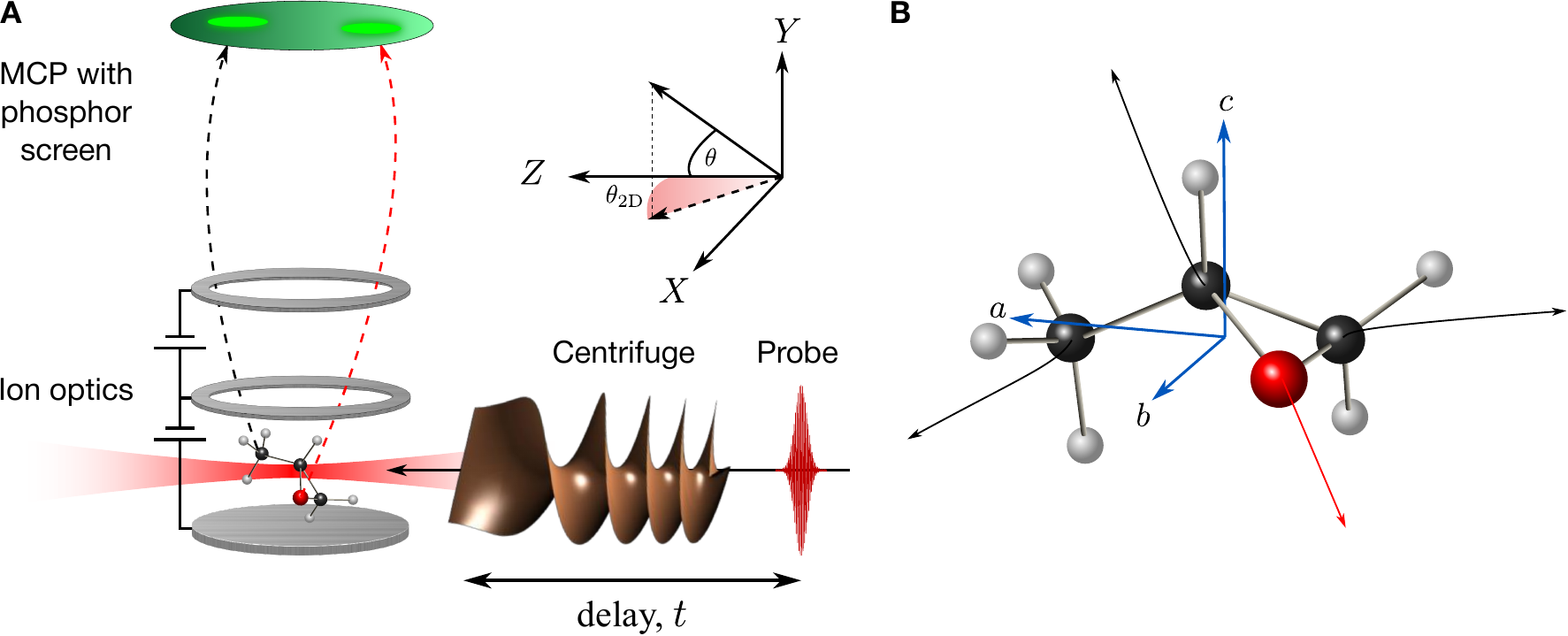} \caption{(\textbf{A}) Schematic illustration of our experimental geometry.
Cold PPO molecules in a seeded helium jet are spun in an optical centrifuge
and Coulomb exploded with a probe pulse between the plates of a conventional
velocity map imaging spectrometer, equipped with a multi-channel plate
(MCP) detector and a phosphor screen. The inset shows the fixed frame
axes and the definition of angles $\theta$ and $\theta_{2\mathrm{D}}$
used in text. (\textbf{B})\textbf{ }Right-handed enantiomer of propylene
oxide, (\emph{R})-PPO, depicted in the frame of its principal axes
$a$, $b$, and $c$ (the corresponding moments of inertia satisfy
$I_{a}<I_{b}<I_{c}$); $a$ axis is close to the principal axis of
polarizability tensor with highest polarization value. Red, black,
and gray spheres represent oxygen, carbon, and hydrogen atoms, respectively.
Coulomb explosion trajectories from a stationary molecule are shown
with thin black and red lines (see Theoretical Methods section for
details).}
\label{fig:FIG1}
\end{figure*}

\subsection*{Experimental Methods and Results}

Our setup for producing the field of an optical centrifuge has been
built according to the original recipe of Karzmarek et al. \cite{Karczmarek1999}
and has been described in an earlier publication \cite{Korobenko2014}.
Briefly, we split the spectrum of broadband laser pulses from a Ti:sapphire
amplifier (10 mJ, 35 fs, repetition rate 1 KHz, central wavelength
795 nm) in two equal parts using a Fourier pulse shaper. The two beams
are first frequency chirped with a chirp rate $\beta$ of equal magnitude
and opposite sign ($\beta=\pm0.3$ THz/ps). The chirped beams are
then circularly polarized with an opposite sense of circular polarization.
Optical interference of these laser fields results in a pulse illustrated
in Fig. \ref{fig:FIG1}A: its polarization vector is rotating in $XY$
plane with an instantaneous angular frequency $\Omega=2\beta t$.
The duration of our centrifuge pulse is 20 ps.

\begin{figure}
\centering{}\includegraphics[width=0.9\columnwidth]{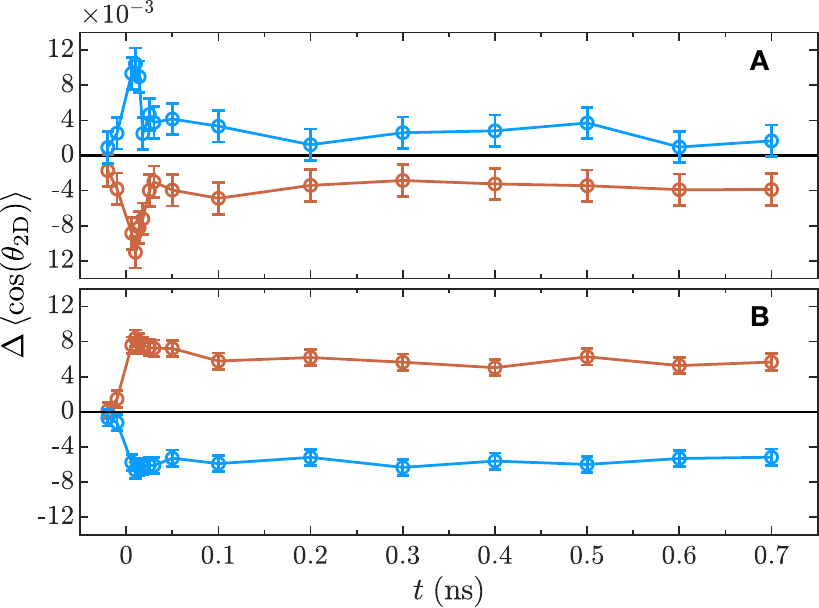} \caption{Experimentally measured 2D orientation factor in the velocity distribution
of (\textbf{A}) $\mathrm{O}^{+}$ and (\textbf{B}) $\mathrm{C}^{+}$
fragments as a function of optical centrifuge-probe pulse delay, $t$.
Orange - right handed molecule, (\emph{R})-PPO; blue - left handed
molecule, (\emph{S})-PPO. Note the reversal of colors between the
two plots.}
\label{fig:FIG2}
\end{figure}

We use a typical velocity map imaging (VMI) setup in which the molecular
jet is intercepted by intense femtosecond probe beam between the plates
of a time-of-flight spectrometer (see Fig. \ref{fig:FIG1}A). Multiple
ionization of a molecule by the probe beam results in Coulomb explosion,
i.e. breaking of the molecule into ionized fragments. As the fragment
ions accelerate towards, and impinge on, the multi-channel plate detector
(MCP), the projection of their velocities on the $XZ$ plane is mapped
on the plane of the detector. Mass selectivity is provided by gating
the MCP at the time of arrival of the fragment of interest. A simplified
model of Coulomb explosion shown in Fig. \ref{fig:FIG1}B predicts
that the velocities of $\mathrm{C}^{+}$ ions (averaged over three
carbon atoms) and $\mathrm{O}^{+}$ ions have opposite projections
on the laboratory $Z$ axis (see Fig. \ref{fig:FIG3}), in full agreement
with the experimental observations (see Fig. \ref{fig:FIG2}).

We extract the information about the molecular orientation in the
laboratory frame as a function of time by recording VMI images at
different centrifuge-probe time delays (see Fig. \ref{fig:FIG1}A).
The experimental observable, bearing the information about the degree
of orientation, is conventionally determined as $\langle\cos(\theta_{2\mathrm{D}})\rangle$.
Here $\theta_{2\mathrm{D}}$ is the angle between the projection of
the fragment's velocity $\mathbf{v}$ on the velocity map imaging
detector plane ($XZ$ plane) and the laboratory $Z$ axis, $\langle\cdots\rangle$
implies averaging over the molecular ensemble. Positive (negative)
values of $\langle\cos(\theta_{2\mathrm{D}})\rangle$ reflect the
orientation of $\mathbf{v}$ along (against) the laboratory $Z$ axis.
In practice, an average of a few million ion fragments were recorded
for each set of experimental conditions, resulting in the precision
of $10^{-3}$ in determining $\langle\cos(\theta_{2\mathrm{D}})\rangle$.
To minimize systematic errors, e.g. due to the inhomogeneous response
of our detector as well as long-term drifts in molecular density and
laser intensity, we define the following quantity (hereafter referred
to as the ``2D orientation factor''):
\[
\Delta\langle\cos(\theta_{2\mathrm{D}})\rangle\equiv\langle\cos(\theta_{2\mathrm{D}})\rangle_{\circlearrowright}-\langle\cos(\theta_{2\mathrm{D}})\rangle_{\circlearrowleft},
\]
where the indices $\circlearrowright$ and $\circlearrowleft$ correspond
to the clockwise and counter-clockwise direction of polarization rotation,
as observed along the laser beam propagation.
In Fig. \ref{fig:FIG2} we plot the experimentally measured $\Delta\langle\cos(\theta_{2\mathrm{D}})\rangle$
for the velocity distributions of $\mathrm{O}^{+}$ and $\mathrm{C}^{+}$
fragments. For both, the enantioselective effect of the centrifuge
is reflected by the opposite sign of the 2D orientation factor for
the two enantiomers. In the case of oxygen ions (Fig. \ref{fig:FIG2}A),
$|\Delta\langle\cos(\theta_{2\mathrm{D}})\rangle|$ reaches the values
of the order of $10^{-2}$ during the interaction with the centrifuge
field (first 20 ps), being positive for left- and negative for right-handed
molecules. When the interaction is over, the degree of orientation
become smaller, but maintain a non-zero value of opposite signs for
at least 700 ps (a maximum accessible delay time in the current experimental
setup). Carbon ions demonstrate similar behavior, shown in Fig. \ref{fig:FIG2}B.
The non-zero 2D orientation factor of $\mathrm{C}^{+}$ also persists
on the full available time scale. As seen, the orientation signals
of $\mathrm{O}^{+}$ and $\mathrm{C}^{+}$ ions are opposite to each
other for both enantiomers, in agreement with theoretical simulations
(see Fig. \ref{fig:FIG3}).

\subsection*{Theoretical Methods and Results}

The chiral molecule is modeled as rigid asymmetric top having anisotropic
polarizability. We carry out both classical and fully quantum simulations
of the laser driven rotational dynamics. For the classical simulations,
the behavior of a thermal ensemble was treated using the Monte Carlo
approach. Our numerical scheme relies on solving the coupled system
of Euler equations and quaternions equations of motion \cite{Art-of-Molecular-Simulation}.
The molecular parameters and details of the scheme may be found in
\cite{Tutunnikov2018}.

For the quantum simulations, we use the symmetric-top wavefunctions
$|JKM\rangle$ as a basis set. Here $J$ is the total angular momentum,
$K$ is its projection on the molecule-fixed $a$ axis, and $M$ is
its projection onto the laboratory fixed $Z$ axis \cite{Zare}. The
interaction potential is expressed in terms of Wigner D-matrices.
Using a unitary transformation, the matrices are transformed to the
asymmetric-top basis. Details of the scheme may be found in \cite{Tutunnikov2019}.

We adopted a simplified model of Coulomb explosion, which assumes
an instantaneous conversion of all constituent atoms into singly charged
ions, while keeping the molecular configuration unchanged during the
interaction with the probe pulse. For finding the asymptotic velocities
of all ions after the Coulomb explosion, we numerically solved the
system of coupled Newton's equations. As the kinetic energy of the
fragments due to Coulomb explosion exceeds the rotational energy,
we neglected the rotational velocities of the atoms at the moment
of explosion. Trajectories of the fragments are shown in Fig. \ref{fig:FIG1}B
by red and black arrowed lines for oxygen and carbon ions, respectively.
A static PPO molecule would eject its ions along these directions
that carry the information about the spatial orientation of the molecule
at the moment of explosion.

\begin{figure}
\centering{}\includegraphics[width=0.9\columnwidth]{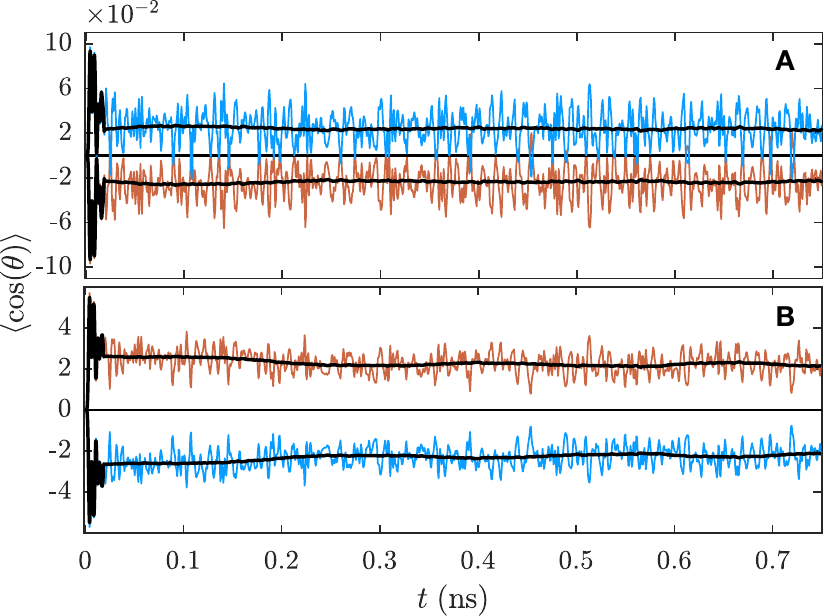} \caption{(\textbf{A}) 3D orientation factor of the oxygen ion velocity. (\textbf{B})
3D orientation factor of the carbon ion velocity (averaged over the
three carbons). Orange - (\emph{R})\textbf{-}PPO; light blue - (\emph{S})-PPO.
Solid black curves are a 100 ps time-averages, $\overline{\langle\cos\left(\theta\right)\rangle}(t)=\frac{1}{\Delta t}\int_{t-\Delta t/2}^{t+\Delta t/2}\langle\cos\left(\theta\right)\rangle\left(t^{\prime}\right)\mathrm{d}t^{\prime}$
of the signals.\textbf{ }Parameters of the optical centrifuge pulse
are: peak intensity $I_{0}=5\times10^{12}\;\mathrm{W/cm^{2}}$, $\beta=0.05\;\mathrm{ps^{-2}}$,
duration $t_{p}=20$ ps. Initial rotation temperature was set to $T=0$
K.}
\label{fig:FIG3}
\end{figure}

Figure \ref{fig:FIG3} shows the results of our simulations, and presents
the conventional (3D) orientation factors, $\braket{\cos(\theta)}$
- the expectation values of the projection of oxygen and carbon ions
velocity vectors on the laser propagation direction ($Z$ axis in
Fig. \ref{fig:FIG1}A). Here $\theta$ is the angle between the velocity
vector and the $Z$ axis, and $\braket{\cdots}$ denotes ensemble
average or quantum expectation value for the classical and quantum
mechanical simulations, respectively. The presented orientation factor
for the carbon ions was obtained by averaging the results for the
three individual molecular carbons. Similar to the experimental results,
the orientation factors of both oxygen and carbon ions are non-zero
long after the end of the optical centrifuge pulse. Note, that these
factors have opposite signs for oxygen and carbon. A detailed analysis
shows the reason for this. When the molecular $a$ axis was trapped
by the optical centrifuge and rotated in the $XY$ plane, the molecule
turned about the $a$ axis such that the middle carbon and oxygen
atoms happened to be on the opposite sides of the $XY$ plane. As
the result of the Coulomb explosion, the two end carbon ions recoil
close to the plane, while the middle carbon and the oxygen ions fly
apart on different sides of the $XY$ plane, resulting in the opposite
orientation factors. The observed permanent orientation is enantioselective
and opposite for the two enantiomers.

\begin{figure}[h]
\centering{}\includegraphics[width=0.9\columnwidth]{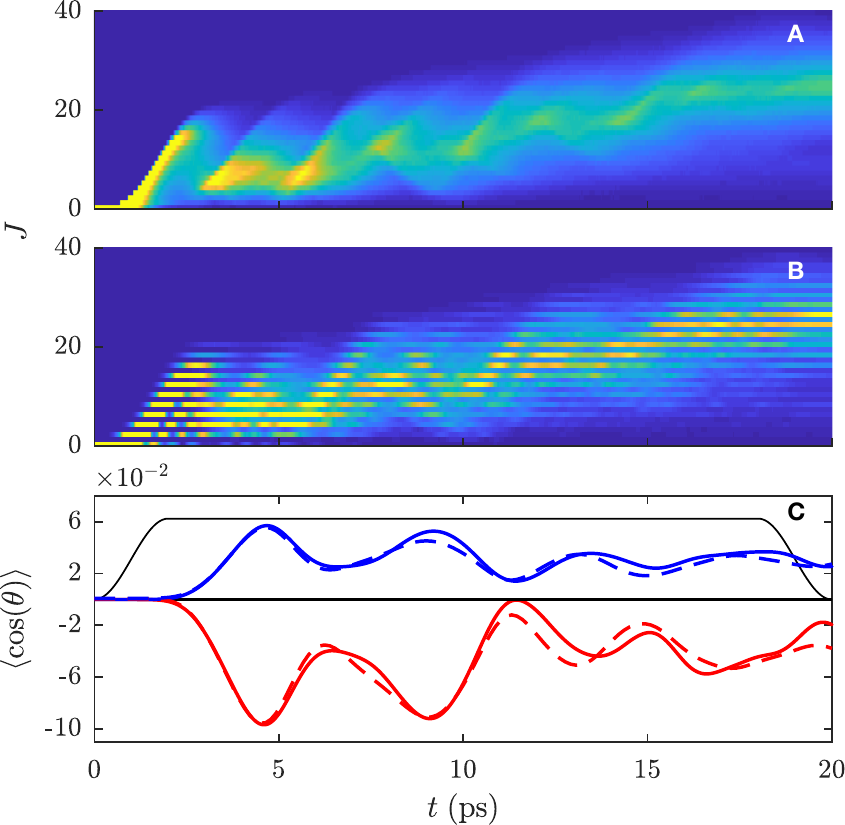} \caption{(\textbf{A},\textbf{B}) Distribution of the angular momentum in units
of $\hbar$, $J$ as a function of time calculated (\textbf{A}) classically
and (\textbf{B}) quantum mechanically. (\textbf{C})\textbf{ }Short
time dynamics of the 3D orientation factors (for (\emph{R})-PPO).
Solid blue and red curves are results of quantum mechanical simulation
for carbon and oxygen ions velocities, respectively, whereas dashed
lines show the results of classical simulations. Solid black line
represents the intensity profile of the optical centrifuge field in
arbitrary units. Parameters of the optical centrifuge are the same
as in Fig. \ref{fig:FIG3}.}
\label{fig:FIG4}
\end{figure}

It is interesting to consider the dynamics of molecules during their
interaction with the optical centrifuge, which eventually leads to
the field-free PESO phenomenon. Figure \ref{fig:FIG4}A,B shows the
distribution of the angular momentum magnitude (in units of $\hbar$)
as a function of time during the optical centrifuge operation (20
ps). It is evident that the molecules undergo a forced angular acceleration.
Figure \ref{fig:FIG4}C shows the simulated (both classically and
quantum mechanically) 3D orientation factors during the first 20 ps
of the evolution. Solid black curve represents the envelope of the
optical centrifuge field showing that the peak intensity is reached
after 2 ps. During this initial stage, molecules are aligned in the
direction of the (barely) rotating polarization, while their orientation
remains zero. The latter rises only after the polarization twist becomes
substantial and induces an orienting torque in the direction of the
aligned most polarizable molecular axis $a$ (see Fig. \ref{fig:FIG1}B).
This supports the orientation mechanism described in \cite{Gershnabel2018,Tutunnikov2018}.
The 3D orientation factors reach their persistent values of $\overline{\langle\cos\left(\theta\right)\rangle}\approx2.5\times10^{-2}$
and $\overline{\langle\cos\left(\theta\right)\rangle}\approx2.3\times10^{-2}$
for oxygen and carbon, respectively, after approximately $12$ ps.
Further acceleration is of little effect on the level of orientation.
For computational reasons, the angular acceleration used in our simulation
is lower than the experimental one (by a factor of 6). The maximum
attainable magnitude of the angular momentum, $J_{\mathrm{max}}$
is proportional to the product $t_{p}\beta$, while the basis size
required for full quantum simulations of asymmetric-top molecule scales
as $J_{\mathrm{max}}^{3}$. The cubic scaling forbids the direct quantum
calculation involving very high angular momentum states, although
some progress is offered by the semi-classical methods \cite{Schmiedt2017}.
Moreover, at high $J$ values, the rigid-top approximation becomes
invalid. However, even with the chosen $\beta$ value, our simulations
qualitatively reproduce the observed effect. Figure \ref{fig:FIG4}
suggests that the degree of molecular orientation does not necessarily
benefit from the acceleration to progressively higher angular velocities,
as only the initial stage of the accelerated spinning contributes
to the desired orientation effect.

\subsection*{Conclusions}

We demonstrated both theoretically and experimentally the persistent
enantioselective orientation (PESO) of chiral molecules induced by
laser fields with twisted polarization. Theoretically, we found very
good agreement between the classical and quantum approaches in the
experimentally relevant range of parameters. The molecular orientation
direction depends on both the sense of polarization twisting and the
handedness of the molecule.

This long-lasting orientation provides new modalities for controlling
the motion of chiral molecules, detecting molecular chirality by means
of nonlinear optics, and potentially for for separating the laser-oriented
enantiomers in external inhomogeneous fields (see e.g. \cite{Gershnabel2011a,Gershnabel2011b,Yachmenev2019},
references therein, and recent reviews \cite{Fleischer2012,Lemeshko2013,Chang2015}).

\subsection*{References and Notes}

\bibliographystyle{unsrt}

\begin{thebibliography}{10}

\bibitem{Cotton}
F.~A. Cotton.
\newblock {\em Chemical Applications of Group Theory}.
\newblock John Wiley \& Sons, Hoboken, NJ, USA, 3\textsuperscript{rd} edition,
  1990.

\bibitem{Koch2018}
C.~P. Koch, M.~Lemeshko, and D.~Sugny.
\newblock Quantum control of molecular rotation.
\newblock {\em arXiv: 1810.11338v1 [quant-ph], Rev. Mod. Phys. in press}, 2019.

\bibitem{Lemeshko2013}
M.~Lemeshko, R.~V. Krems, J.~M. Doyle, and S.~Kais.
\newblock Manipulation of molecules with electromagnetic fields.
\newblock {\em Mol. Phys.}, 111(12-13):1648--1682, 2013.

\bibitem{Fleischer2012}
S.~Fleischer, Y.~Khodorkovsky, E.~Gershnabel, Y.~Prior, and I.~Sh. Averbukh.
\newblock Molecular alignment induced by ultrashort laser pulses and its impact
  on molecular motion.
\newblock {\em Isr. J. Chem.}, 52(5):414--437, 2012.

\bibitem{Ohshima2010}
Y.~Ohshima and H.~Hasegawa.
\newblock Coherent rotational excitation by intense nonresonant laser fields.
\newblock {\em Int. Rev. Phys. Chem.}, 29(4):619--663, 2010.

\bibitem{Stapelfeldt2003}
H.~Stapelfeldt and T.~Seideman.
\newblock Colloquium: Aligning molecules with strong laser pulses.
\newblock {\em Rev. Mod. Phys.}, 75:543--557, Apr 2003.

\bibitem{Lin2018}
K.~Lin, I.~Tutunnikov, J.~Qiang, J.~Ma, Q.~Song, Q.~Ji, W.~Zhang, H.~Li,
  F.~Sun, X.~Gong, H.~Li, P.~Lu, H.~Zeng, Y.~Prior, I.~Sh. Averbukh, and J.~Wu.
\newblock All-optical field-free three-dimensional orientation of
  asymmetric-top molecules.
\newblock {\em Nat. Commun.}, 9(1):5134, 2018.

\bibitem{Yachmenev2016}
A.~Yachmenev and S.~N. Yurchenko.
\newblock Detecting chirality in molecules by linearly polarized laser fields.
\newblock {\em Phys. Rev. Lett.}, 117:033001, 2016.

\bibitem{Gershnabel2018}
E.~Gershnabel and I.~Sh. Averbukh.
\newblock Orienting asymmetric molecules by laser fields with twisted
  polarization.
\newblock {\em Phys. Rev. Lett.}, 120:083204, 2018.

\bibitem{Tutunnikov2018}
I.~Tutunnikov, E.~Gershnabel, S.~Gold, and I.~Sh. Averbukh.
\newblock Selective orientation of chiral molecules by laser fields with
  twisted polarization.
\newblock {\em J. Phys. Chem. Lett}, 9(5):1105--1111, 2018.

\bibitem{Milner2019}
A.~A. Milner, J.~A.~M. Fordyce, I.~MacPhail-Bartley, W.~Wasserman, V.~Milner,
  I.~Tutunnikov, and I.~Sh. Averbukh.
\newblock Controlled enantioselective orientation of chiral molecules with an
  optical centrifuge.
\newblock {\em Phys. Rev. Lett.}, 122:223201, 2019.

\bibitem{Harde1991}
H.~Harde, S.~Keiding, and D.~Grischkowsky.
\newblock {THz} commensurate echoes: Periodic rephasing of molecular
  transitions in free-induction decay.
\newblock {\em Phys. Rev. Lett.}, 66:1834--1837, 1991.

\bibitem{Dion1999}
C.~M. Dion, A.~D. Bandrauk, O.~Atabek, A.~Keller, H.~Umeda, and Y.~Fujimura.
\newblock Two-frequency {IR} laser orientation of polar molecules. {N}umerical
  simulations for {HCN}.
\newblock {\em Chem. Phys. Lett.}, 302(3-4):215--223, 1999.

\bibitem{Averbukh2001}
I.~Sh. Averbukh and R.~Arvieu.
\newblock Angular focusing, squeezing, and rainbow formation in a strongly
  driven quantum rotor.
\newblock {\em Phys. Rev. Lett.}, 87:163601, 2001.

\bibitem{Machholm2001}
M.~Machholm and N.~E. Henriksen.
\newblock Field-free orientation of molecules.
\newblock {\em Phys. Rev. Lett.}, 87:193001, 2001.

\bibitem{Fleischer2011}
S.~Fleischer, Y.~Zhou, R.~W. Field, and K.~A. Nelson.
\newblock Molecular orientation and alignment by intense single-cycle {THz}
  pulses.
\newblock {\em Phys. Rev. Lett.}, 107:163603, 2011.

\bibitem{Kitano2013}
K.~Kitano, N.~Ishii, N.~Kanda, Y.~Matsumoto, T.~Kanai, M.~Kuwata-Gonokami, and
  J.~Itatani.
\newblock Orientation of jet-cooled polar molecules with an intense
  single-cycle {THz} pulse.
\newblock {\em Phys. Rev. A}, 88:061405, 2013.

\bibitem{Damari2016}
R.~Damari, S.~Kallush, and S.~Fleischer.
\newblock Rotational control of asymmetric molecules: Dipole- versus
  polarizability-driven rotational dynamics.
\newblock {\em Phys. Rev. Lett.}, 117:103001, 2016.

\bibitem{Daems2005}
D.~Daems, S.~Gu\'erin, D.~Sugny, and H.~R. Jauslin.
\newblock Efficient and long-lived field-free orientation of molecules by a
  single hybrid short pulse.
\newblock {\em Phys. Rev. Lett.}, 94:153003, 2005.

\bibitem{Gershnabel2006}
E.~Gershnabel, I.~Sh. Averbukh, and R.~J. Gordon.
\newblock Orientation of molecules via laser-induced antialignment.
\newblock {\em Phys. Rev. A}, 73:061401, 2006.

\bibitem{Egodapitiya2014}
K.~N. Egodapitiya, S.~Li, and R.~R. Jones.
\newblock Terahertz-induced field-free orientation of rotationally excited
  molecules.
\newblock {\em Phys. Rev. Lett.}, 112:103002, 2014.

\bibitem{Vrakking1997}
M.~J.~J. Vrakking and S.~Stolte.
\newblock Coherent control of molecular orientation.
\newblock {\em Chem. Phys. Lett.}, 271(4-6):209--215, 1997.

\bibitem{Kanai2001}
T.~Kanai and H.~Sakai.
\newblock Numerical simulations of molecular orientation using strong,
  nonresonant, two-color laser fields.
\newblock {\em J. Chem. Phys}, 115(12):5492--5497, 2001.

\bibitem{De2009}
S.~De, I.~Znakovskaya, D.~Ray, F.~Anis, Nora~G. Johnson, I.~A. Bocharova,
  M.~Magrakvelidze, B.~D. Esry, C.~L. Cocke, I.~V. Litvinyuk, and M.~F. Kling.
\newblock Field-free orientation of {CO} molecules by femtosecond two-color
  laser fields.
\newblock {\em Phys. Rev. Lett.}, 103:153002, 2009.

\bibitem{Oda2010}
K.~Oda, M.~Hita, S.~Minemoto, and H.~Sakai.
\newblock All-optical molecular orientation.
\newblock {\em Phys. Rev. Lett.}, 104:213901, 2010.

\bibitem{JW2010}
J.~Wu and H.~Zeng.
\newblock Field-free molecular orientation control by two ultrashort dual-color
  laser pulses.
\newblock {\em Phys. Rev. A}, 81:053401, 2010.

\bibitem{Frumker2012}
E.~Frumker, C.~T. Hebeisen, N.~Kajumba, J.~B. Bertrand, H.~J. W\"orner,
  M.~Spanner, D.~M. Villeneuve, A.~Naumov, and P.~B. Corkum.
\newblock Oriented rotational wave-packet dynamics studies via high harmonic
  generation.
\newblock {\em Phys. Rev. Lett.}, 109:113901, 2012.

\bibitem{Takemoto2008}
N.~Takemoto and K.~Yamanouchi.
\newblock Fixing chiral molecules in space by intense two-color phase-locked
  laser fields.
\newblock {\em Chem. Phys. Lett.}, 451(1):1 -- 7, 2008.

\bibitem{Tutunnikov2019}
I.~Tutunnikov, J.~Flo{\ss}, E.~Gershnabel, P.~Brumer, and I.~Sh. Averbukh.
\newblock Laser induced persistent orientation of chiral molecules.
\newblock {\em arXiv: 1905.12609 [physics.chem-ph]}, 2019.

\bibitem{Fleischer2009}
S.~Fleischer, Y.~Khodorkovsky, Y.~Prior, and I.~Sh. Averbukh.
\newblock Controlling the sense of molecular rotation.
\newblock {\em New J. Phys.}, 11(10):105039, 2009.

\bibitem{Kitano2009}
K.~Kitano, H.~Hasegawa, and Y.~Ohshima.
\newblock Ultrafast angular momentum orientation by linearly polarized laser
  fields.
\newblock {\em Phys. Rev. Lett.}, 103:223002, 2009.

\bibitem{Zhdanovich2011}
S.~Zhdanovich, A.~A. Milner, C.~Bloomquist, J.~Flo\ss{}, I.~Sh. Averbukh, J.~W.
  Hepburn, and V.~Milner.
\newblock Control of molecular rotation with a chiral train of ultrashort
  pulses.
\newblock {\em Phys. Rev. Lett.}, 107:243004, 2011.

\bibitem{Bloomquist2012}
C.~Bloomquist, S.~Zhdanovich, A.~A. Milner, and V.~Milner.
\newblock Directional spinning of molecules with sequences of femtosecond
  pulses.
\newblock {\em Phys. Rev. A}, 86:063413, 2012.

\bibitem{Floss2012}
J.~Flo{\ss} and I.~Sh. Averbukh.
\newblock Molecular spinning by a chiral train of short laser pulses.
\newblock {\em Phys. Rev. A}, 86:063414, 2012.

\bibitem{Karras2015}
G.~Karras, M.~Ndong, E.~Hertz, D.~Sugny, F.~Billard, B.~Lavorel, and
  O.~Faucher.
\newblock Polarization shaping for unidirectional rotational motion of
  molecules.
\newblock {\em Phys. Rev. Lett.}, 114:103001, Mar 2015.

\bibitem{Prost2017}
E.~Prost, H.~Zhang, E.~Hertz, F.~Billard, B.~Lavorel, P.~Bejot, J.~Zyss, I.~Sh.
  Averbukh, and O.~Faucher.
\newblock Third-order-harmonic generation in coherently spinning molecules.
\newblock {\em Phys. Rev. A}, 96:043418, 2017.

\bibitem{Prost2018}
E.~Prost, E.~Hertz, F.~Billard, B.~Lavorel, and O.~Faucher.
\newblock Polarization-based tachometer for measuring spinning rotors.
\newblock {\em Opt. Express}, 26(24):31839--31849, 2018.

\bibitem{Karczmarek1999}
J.~Karczmarek, J.~Wright, P.~Corkum, and M.~Ivanov.
\newblock Optical centrifuge for molecules.
\newblock {\em Phys. Rev. Lett.}, 82:3420--3423, 1999.

\bibitem{Villeneuve2000}
D.~M. Villeneuve, S.~A. Aseyev, P.~Dietrich, M.~Spanner, M.~Yu. Ivanov, and
  P.~B. Corkum.
\newblock Forced molecular rotation in an optical centrifuge.
\newblock {\em Phys. Rev. Lett.}, 85:542--545, 2000.

\bibitem{Yuan2011}
L.~Yuan, S.~W. Teitelbaum, A.~Robinson, and A.~S. Mullin.
\newblock Dynamics of molecules in extreme rotational states.
\newblock {\em Proc. Natl. Acad. Sci. U. S. A.}, 108(17):6872--6877, 2011.

\bibitem{Korobenko2014-PRL}
A.~Korobenko, A.~A. Milner, and V.~Milner.
\newblock Direct observation, study, and control of molecular superrotors.
\newblock {\em Phys. Rev. Lett.}, 112:113004, 2014.

\bibitem{Korobenko2018}
A.~Korobenko.
\newblock Control of molecular rotation with an optical centrifuge.
\newblock {\em J. Phys. B}, 51(20):203001, 2018.

\bibitem{Korobenko2014}
A.~Korobenko, A.~A. Milner, J.~W. Hepburn, and V.~Milner.
\newblock Rotational spectroscopy with an optical centrifuge.
\newblock {\em Phys. Chem. Chem. Phys.}, 16:4071--4076, 2014.

\bibitem{Art-of-Molecular-Simulation}
D.~C. Rapaport.
\newblock {\em The Art of Molecular Dynamics Simulation}.
\newblock Cambridge University Press, 2\textsuperscript{nd} edition, 2004.

\bibitem{Zare}
R.~Zare.
\newblock {\em Angular momentum : understanding spatial aspects in chemistry
  and physics}.
\newblock Wiley, New York, 1988.

\bibitem{Schmiedt2017}
H.~Schmiedt, S.~Schlemmer, S.~N. Yurchenko, A.~Yachmenev, and P.~Jensen.
\newblock A semi-classical approach to the calculation of highly excited
  rotational energies for asymmetric-top molecules.
\newblock {\em Phys. Chem. Chem. Phys.}, 19:1847--1856, 2017.

\bibitem{Gershnabel2011a}
E.~Gershnabel and I.~Sh. Averbukh.
\newblock Electric deflection of rotating molecules.
\newblock {\em J. Chem. Phys.}, 134(5):054304, 2011.

\bibitem{Gershnabel2011b}
E.~Gershnabel and I.~Sh. Averbukh.
\newblock Deflection of rotating symmetric top molecules by inhomogeneous
  fields.
\newblock {\em J. Chem. Phys.}, 135(8):084307, 2011.

\bibitem{Yachmenev2019}
A.~Yachmenev, J.~Onvlee, E.~Zak, A.~Owens, and J.~K{\"u}pper.
\newblock Field-induced diastereomers for chiral separation.
\newblock {\em arXiv: 1905.07166 [physics.chem-ph]}, 2019.

\bibitem{Chang2015}
Y.-P. Chang, D.~A. Horke, S.~Trippel, and J.~K{\"u}pper.
\newblock Spatially-controlled complex molecules and their applications.
\newblock {\em Int. Rev. Phys. Chem.}, 34(4):557--590, 2015.

\end{thebibliography}

\noindent \begin{flushleft}
\textbf{Acknowledgments:} This work was supported by the Israel Science
Foundation (Grant No. 746/15), the ISF-NSFC joint research program
(Grant No. 2520/17), and by Natural Sciences and Engineering Research
Council of Canada grants to P.B. and V.M. The work of A.A.M. and V.M.
was carried out under the auspices of the Canadian Center for Chirality
Research on Origins and Separation (CHIROS), funded by Canada Foundation
for Innovation. I.A. acknowledges support as the Patricia Elman Bildner
Professorial Chair, and thanks the UBC Department of Physics \& Astronomy
for hospitality extended to him during his sabbatical stay. This research
was made possible in part by the historic generosity of the Harold
Perlman Family. \textbf{Author contributions:} I.T., I.A., P.B., and
V.M. initiated the study. A.A.M. and V.M. designed and carried out
the experiment. I.T., J.F., E.G. performed the simulations. I.A.,
P.B., and V.M. supervised and guided the work. All authors contributed
to the data analysis and writing the manuscript. \textbf{Competing
interests:} The authors declare that they have no competing interests.
\par\end{flushleft}
\end{document}